%% file: mimicry_arXiv.tex
\documentclass[aps,prd,twocolumn,preprintnumbers,groupedaddress,nofootinbib]{revtex4}

\usepackage[latin1]{inputenc}
\usepackage[english]{babel}
\usepackage{amsmath}    
\usepackage{amssymb}
\usepackage{amsthm}
\usepackage{upgreek}
\usepackage{bbm}        
\usepackage[pdftex]{color}
\usepackage{wasysym}
\usepackage{slashed}
\usepackage{multirow, booktabs}
\usepackage{graphicx}
\usepackage{afterpage}
\usepackage{exscale}
\usepackage[pdftex,colorlinks,pdfpagelabels]{hyperref}
\usepackage[figure,table]{hypcap} 
\hypersetup{
   bookmarksnumbered,
   pdfstartview={FitV},
   pdfpagemode={UseOutlines},
   pdfauthor={Jasper Hasenkamp},
   pdftitle={Daughters mimic sterile neutrinos (almost!) perfectly}, 
   pdfsubject={scientific publication}
   ,citecolor={blue},
   linkcolor={blue},
   urlcolor={blue},
   filecolor={blue}
} 

\input{commands.tex}

\begin{document}
 
\date{\mbox{26th May 2014}}

\title{ 
\bf 
Daughters mimic sterile neutrinos (almost!) perfectly
}
\author{Jasper Hasenkamp}
\email{Jasper.Hasenkamp@nyu.edu}
\affiliation{\mbox{CCPP, Department of Physics, New York University, New York, NY 10003, USA}}

\begin{abstract}
\noindent
Since only recently, cosmological observations are sensitive to hot dark matter (HDM) admixtures with sub-eV mass, $\mnuseff < \eV$, that are not fully-thermalised, $\D\Neff< 1$.
We argue that their almost automatic interpretation as a sterile neutrino species is neither from theoretical nor practical parsimony principles preferred over HDM formed by decay products (daughters) of an out-of-equilibrium particle decay.
While daughters mimic sterile neutrinos in $\Neff$ and $\mnuseff$, 
there are opportunities to assess this possibility in likelihood analyses. Connecting cosmological parameters and moments of momentum distribution functions, we show that 
--also in the case of mass-degenerate daughters with indistinguishable main physical effects--
the mimicry breaks down when the next moment, the skewness, is considered.  Predicted differences of order one in the root-mean-squares of absolute momenta are too small for current sensitivities. 
\end{abstract}
\maketitle

\section{Introduction}
We are in the era of precision cosmology. By observing the cosmic microwave background (CMB), the Planck satellite provided data that allows us to determine base quantities in the standard model of cosmology ($\L$CDM) like the energy density of cold dark matter (CDM) on the percent level~\cite{Ade:2013zuv}.
Especially, if combined with further cosmological observations, most violations of assumptions and deviations from predictions are constrained severely. 

We focus on recent hints for a hot dark matter (HDM) \textit{admixture}~\cite{Wyman:2013lza,Hamann:2013iba,Battye:2013xqa} 
parametrised by an effective number of additional neutrino species $\D\Neff$ and effective HDM mass of~\cite{Hamann:2013iba}
\begin{equation}
 \label{hdmsignal}
\D\Neff = 0.61\pm0.30  \text{,} \quad m_\text{hdm}^\text{eff} = (0.41 \pm 0.13) \eV \, .
\end{equation}
Only recently, precisions became high enough to possibly find evidence for such a \textit{sub-eV, not fully-thermalised} species. Consequently, it is of utmost importance to scrutinise  and improve the accuracy of the considered observations, see for example~\cite{Costanzi:2013bha,Efstathiou:2013via,Paranjape:2014lga,Leistedt:2014sia}. For the same reason, the time to consider such HDM admixtures independent of the exact confidence intervals is now.

The common particle physics interpretation is an additional, uncharged (= sterile) neutrino $\nus$ species that mixes with the active neutrinos and, consequently, is produced when these scatter in the early universe.
The beauty of this interpretation is its parsimony: Firstly, there are hints from neutrino oscillation experiments for sterile neutrinos with \order{\text{eV}} masses and mixings that thermalise them. No further physics needs to be introduced, actually, from symmetry arguments one might expect even three such particles.  Secondly, the cosmological model is amended by only one free parameter per particle, the sterile neutrino's mass, since full-thermalisation fixes its temperature to the neutrino temperature and thus implies $\D\Neff=1$.
However, the cosmological signal just does not fit that interpretation as illustrated in Fig.~\ref{fig:posterior}.

\begin{figure}%
\centering
 \includegraphics[width=0.8\columnwidth]{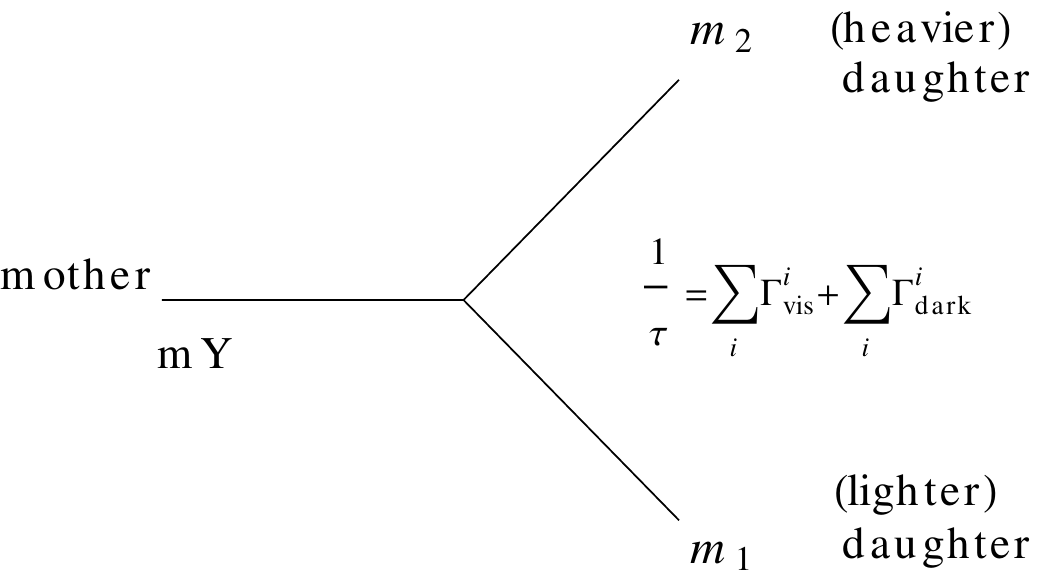}
\caption{Our nomenclature for a cosmological particle decay and the corresponding degrees of freedom.
The, in principle, arbitrarily large number is argued to reduce to three, which is further reduced to only two actually independent degrees of freedom by a physical parameter degeneracy.}
\label{fig:dof}
\end{figure}
Accepting the possibility of $\D\Neff < 1$, the cosmological model is amended by a second free parameter.
Attempts to reconcile sterile neutrinos with cosmology include the addition of light, interacting particles~\cite{Steigman:2013yua} and new neutrino interactions~\cite{Hannestad:2013ana,Dasgupta:2013zpn} that, indeed, lead to HDM improving the cosmological fit if they are gauged and shared by CDM~\cite{Bringmann:2013vra}. Along the way, these ideas also give up the first parsimony argument. 
At that state we find it to be indicated to reconsider the interpretation of~\eqref{hdmsignal} being due to a thermally produced sterile neutrino (\nus HDM), even if for the sake of identifying the limits of our understanding.
In this work, we consider the possibility that hot dark matter is formed by the decay products of an out-of-equilibrium particle decay (dpHDM). So it originates from particle decay instead of scattering. 
Cosmological particle  decays with in part dramatic consequences are naturally expected in many theories beyond the Standard Model of particle physics (SM). They do not only occur naturally, if gravity with the prominent gravitino and Polonyi problems is considered, see~\cite{Hasenkamp:2011xh,Hasenkamp:2011em,Hasenkamp:2012ii,Bae:2013qr,Graf:2013xpe,Higaki:2013vuv,DiBari:2013dna,Conlon:2013isa,Park:2013bza,Hooper:2013nia,Kelso:2013nwa,Jeong:2013oza} and references therein.
It is remarkable that dpHDM does not require the introduction of any small mass scales as \nus HDM. 
From a particle physical point of view a decay, as depicted schematically in Fig.~\ref{fig:dof}, might be described by an arbitrarily large number of degrees of freedom just like a ``dark sector'' with new interactions. As in that case the HDM component is, nevertheless, described by only two actually independent degrees of freedom. So from an observational point of view the complexity of the cosmological models is the same.

We will explore, which scenarios of dpHDM have for what reason the potential to be distinguished from \nus HDM and in which case dpHDM
is indistinguishable from \nus HDM in analyses like~\cite{Hamann:2013iba} utilising CMB and large-scale structure observations.
In any case, the cosmological impact of dpHDM cannot be identical, because dpHDM possesses a different momentum distribution function than \nus HDM. We will show at which point and how \textit{any mimicry must break down}.

This work is organised as follows:
In the next section we list main physical effects, define parameters and remind the case of \nus HDM, where we also assign briefly the current observational evidence.
In Sec.~\ref{sec:mimicry} we show how decay products mimic sterile neutrinos, translate observations, point out in which cases there are testable differences in the main physical effects, and how any mimicry must break down in the third cosmological parameter.
In Sec.~\ref{sec:conclusions} we summarise and conclude.
\section{Preliminaries}
\label{sec:preliminaries}
{\it \bf Physical effects (neutrino case) --}
A population of free-streaming particles, which becomes non-relativistic
after photon decoupling, affects the cosmological background and the evolution of perturbations~\cite{Lesgourgues:2006nd}.
Its main physical effects are related to:

1) its contribution to the radiation energy density $\rhorad$ of the Universe before photon decoupling.
 Given
in terms of an effective number of neutrino species $\Neff$ the radiation energy density reads
\begin{equation}
\label{rhorad}
  \r_\text{rad} = \left( 1 + \Neff \frac{7}{8} \left(\frac{T_\n}{T_\g}\right)^4 \right) \r_\g \, ,
 \end{equation}
such that it is split into a sum of the
energy density in photons $\r_\g=(\pi^2/15) T^4$ and the relativistic energy density in anything else.
The Standard Model of particle physics (SM) contains three active neutrinos with $\Neff^\text{sm}=3.046$~\cite{Mangano:2005cc}
 and temperature ratio $T_\n/T_\g = (4/11)^{1/3}$.
The small correction in $\Neff^\text{sm}$ is due to incomplete neutrino decoupling at $e^+e^-$-annihilation.
Any departure from the standard scenario, which increases the expansion rate of the Universe and could shift the time of matter-radiation equality, is then parametrised as a summand in $\Neff = \Neff^\text{sm} + \D\Neff$,
such that the active neutrinos correspond to $\D\Neff=0$ by construction.

2) its non-relativistic energy density today $\r_\text{hdm}^0=n_\text{hdm}^0 m_\text{hdm}$,
which  is given by the mass $m_\text{hdm}$ and today's number density $n_\text{hdm}^0$ of the population.
Free-streaming particles do not cluster on scales below the free-streaming scale and thus damp fluctuations.
The extent of the arising amplitude reduction in the matter power spectrum due to the free-streaming population is $\simeq 8 \,\O_\text{hdm}/\O_\text{m}$~\cite{Lesgourgues:2006nd}, so that the HDM fraction $f=\O_\text{hdm}/\O_\text{m}$ provides a useful parametrisation. Note that such a small HDM \textit{admixture} leads to a step-like feature and not, for example, to a cut-off.

Assuming the active neutrinos are degenerate in mass, their today's energy density normalised by the critical energy density $\rhoc$ reads
\begin{equation}
\label{rhonu}
 \O_\n^0 = 
\frac{\Neff^\text{sm}}{11} \frac{n_\g^0}{\rhoc} \sumnu  \Leftrightarrow \O_\n^0 h^2  \simeq 0.0108 \frac{\sumnu}{\text{eV}} 
\end{equation}
with today's number density of CMB photons given by $n_\g^0 = (2 \z(3)/\pi^2) T_0^3$,
if $T_0=2.7255$ K denotes the CMB temperature today and $\z$ the zeta function. 
By construction~\eqref{rhonu} corresponds to the $\D\Neff = 0$ case.
In a $\L$CDM model amended to include massive (degenerate) active neutrinos this sum of neutrino masses $\sumnu$ is usually ``observed,''
when cosmological parameters are determined.
If $\sumnu$ is not freed in an analysis (``pure'' $\L$CDM), a minimal-mass normal hierarchy can be assumed, which is accurately approximated for current cosmological data as a single massive eigenstate with $m_\nu = 0.06 \eV$~\cite{Ade:2013zuv}.

3) the minimum of the comoving free-streaming wavenumber $k^\text{nr}$. 
This is the scale at which the suppression of fluctuations in the matter power spectrum sets in. It is given by the time of the transition when the population becomes non-relativistic~\cite{Lesgourgues:2006nd}.
In Appendix~\ref{appendix:A} we make this point explicit arriving at
\begin{equation}
\label{knr}
 k^\text{nr} \simeq 4.08 \times 10^{-4} \, \O_\text{m}^\frac{1}{2} \fb{\Tnr}{T_0}{\frac{1}{2}} h \Mpc^{-1} ,
\end{equation}
which depends on the properties of the free-streaming population via the temperature of the Universe at transition $\Tnr$ only.
It is defined by
\begin{equation}
\label{defTnr}
 \langle p \rangle(\Tnr) = m \, ,
\end{equation}
where the angle brackets indicate the average of the particles absolute momenta, $p=|\vec p|$.
If we kept the directional information, the population is on average at rest due to the isotropy of the Universe, $\langle \vec p \rangle =0$.
The root-mean-square momentum of the population equals the mean of its distribution of absolute momenta, $\langle \vec p \rangle_\text{rms} = \langle p \rangle$, which is sometimes just called ''mean of the distribution``.
For degenerate neutrinos~\eqref{defTnr} reads
\begin{equation}
\label{Tnrnu1}
\langle p_\nu \rangle (\Tnr_\nu) 
= \frac{7 \pi^4}{180 \z(3)} \left(\frac{4}{11}\right)^\frac{1}{3} \Tnr_\nu
= \frac{1}{3} \sumnu 
\end{equation}
giving
\begin{equation}
\label{Tnrnu2}
 \Tnr_\nu = \frac{1}{3} \fb{11}{4}{\frac{1}{3}} \frac{180 \z(3)}{7 \pi^4}  \sumnu \simeq 0.148 \sumnu \, .  
\end{equation}
 We will use $\Tnr$ later on when comparing different origins of HDM.
Insertion in~\eqref{knr} yields the expected result
\begin{equation}
 k^\text{nr}_\nu \simeq 0.0103 \, \O_\text{m}^\frac{1}{2}  \fb{\sumnu}{\text{eV}}{\frac{1}{2}} h \Mpc^{-1} \, .
\end{equation}
The corresponding free-streaming scale evaluated today reads
\begin{equation}
 \l^\text{fs}_\nu \simeq  24 \, \frac{\text{eV}}{\sumnu} h^{-1} \Mpc .
\end{equation}
Altogether, we see that $\sumnu$ sets both $\O_\nu^0$ (thus $f_\nu$) and $\Tnr_\nu$ (thus $k^\text{nr}_\nu$). At the same time,
the neutrino temperature is fixed by SM weak interactions.

{\bf Sterile neutrino (and current evidence) --}
\begin{figure}
\includegraphics[width=\columnwidth]{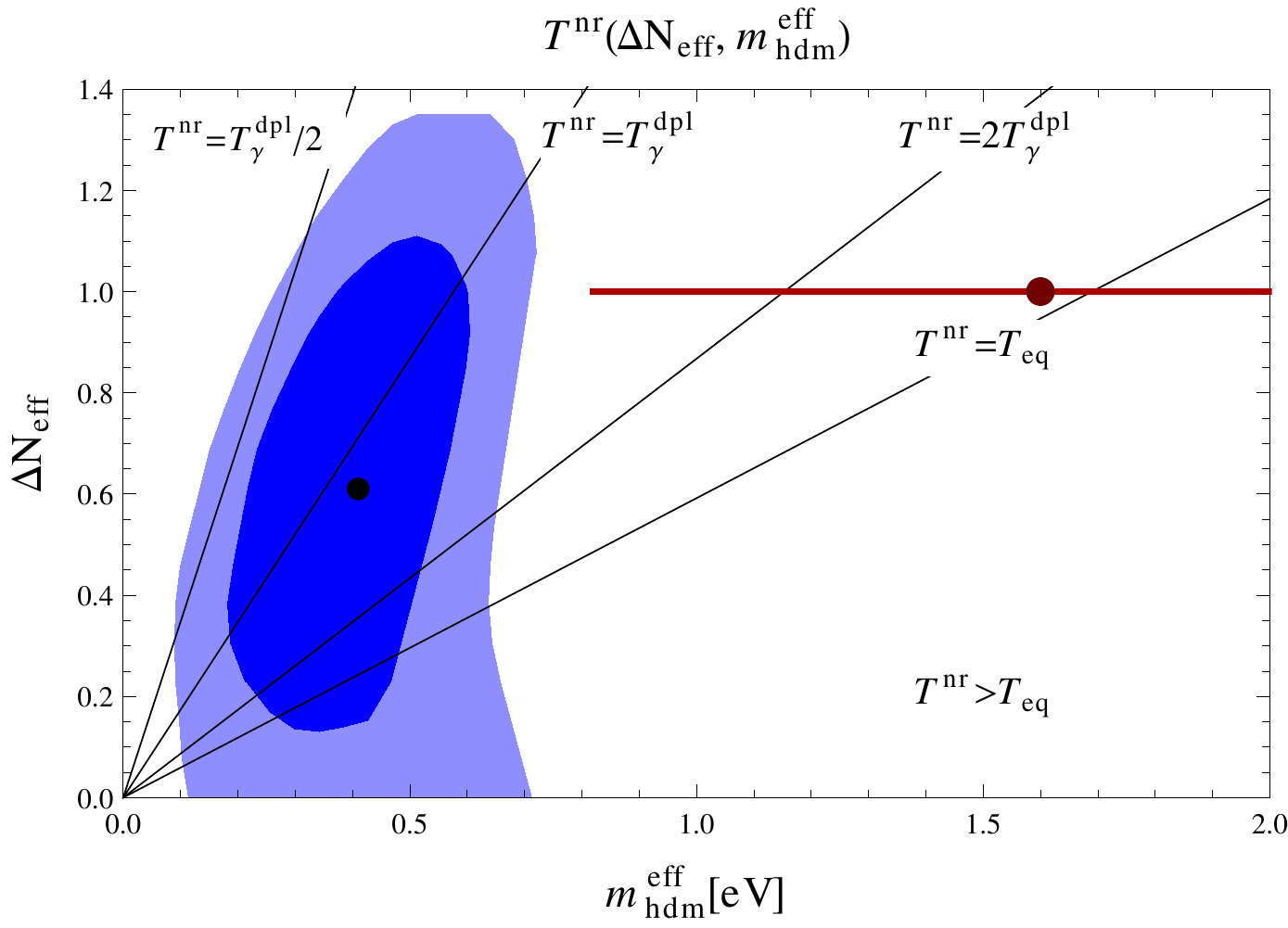} 
\caption{Joint 68\%- and 95\%-credible contours of the marginalised posterior  found in~\cite{Hamann:2013iba} for $\D\Neff$ and $\mnuseff$ of a thermal sterile neutrino model with their 1-d means marked by a dot. Also marked is a best-fit value and the corresponding 3-$\s$ range from oscillation anomalies~\cite{Giunti:2013aea} in a model with one sterile neutrino. The 1+3+1 mass scheme with two sterile neutrinos, which actually might be preferred depending on the datasets considered, implies a large minimal mass like $3.2\eV$~\cite{Kopp:2013vaa}. 
}
\label{fig:posterior}
\end{figure}
As prime example for hot thermal relics we consider a light sterile neutrino $\nu_s$ that is
thermalised but with a different temperature than the active neutrinos (\nus HDM).\footnote{
In other scenarios sterile neutrinos might be produced non-thermally. As we are going to compare the decay case with thermal production, we do not include such possibilities and consider the sterile neutrino to possess a Fermi-Dirac distribution, see also Sec.~\ref{sec:breakdown}.
}
In this case the parameters under consideration are given by:

1) If $T_{\nu_s}$ denotes the temperature of the sterile neutrino population,
\begin{equation}
\label{DNeffnus}
 \D\Neff = \left(\frac{T_{\nu_s}}{T_\nu}\right)^4 \, .
\end{equation}
Inspecting~\eqref{rhorad} we see that~\eqref{DNeffnus} holds by construction for a fermion. Thus a fully-thermalised neutrino species with the energy density $\r_{1\n}=(7/8) (4/11)^{4/3} \r_\g$ corresponds to $\D\Neff=1$.
Other particle natures can be considered by appropriate
 factors in~\eqref{DNeffnus}, for example, $4/7$ for a Nambu-Goldstone boson
or $8/7$ for a massless $U(1)$ gauge boson.
If the sterile neutrino shared the bath temperature once, but decoupled earlier than the active neutrinos,
there is a one-to-one correspondence between its decoupling  temperature $T_{\nu_s}^\text{dpl}$ and its temperature
at later times. It is colder as it missed heating compared  to the bath, when
particles annihilated away. So $\D\Neff = (\gast(T_{\nu}^\text{dpl})/\gast(T_{\nu_s}^\text{dpl}))^{4/3}$,
where $\gast$ is the effective number of relativistic degrees of freedom and the decoupling temperature depends on the sterile neutrino's couplings to the bath.

The evidence in~\cite{Hamann:2013iba} for $\D\Neff > 0$ is mainly driven by local measurements of $H_0$ and seem to be supported by lensing observations~\cite{Hamann:2013iba,Battye:2013xqa}.
Independent evidence arises from the large tensor-to-scalar ratio reported by the BICEP collaboration~\cite{Ade:2014xna}.
In the simplest model of inflation a large ratio implies a large scalar spectral index that increases the power, in particular, at higher multipoles.
Whereas an increased $\Neff$ suppresses power at higher multipoles~\cite{Hou:2012xq}, which is due to increased Silk damping caused by the increased expansion rate~\cite{Hou:2011ec}.

2) The sterile neutrino's non-relativistic energy density today,
\begin{equation}
\label{Onus}
\O_{\nu_s}^0 = \frac{3}{11} \frac{n_\g^0}{\rhoc}  \fb{T_{\nu_s}}{T_\nu}{3}  \frac{\mnus}{\text{eV}} 
\Leftrightarrow \O_{\nu_s}^0 h^2 \simeq 0.0106 \frac{\mnuseff}{\text{eV}}  \, ,
\end{equation}
can be understood by comparison with~\eqref{rhonu}.
The case of interest is one sterile neutrino, which is much heavier than the active neutrinos. Thus
the sum of neutrino masses is to be replaced by the sterile neutrino mass $\mnus$. 
The active neutrinos are then taken into account separately as a single massive eigenstate with $m_\nu = 0.06 \eV$.
Number densities decrease as $\propto T^3$, which explains the temperature ratio to the third power.
We just defined the effective hot dark matter mass 
\begin{equation}
\mnuseff = \mnus (\Tnus/T_\n)^{1/3} .
\end{equation}
Other particle natures can be considered by appropriate
 factors in~\eqref{Onus}, for example, $2/3$ for a Nambu-Goldstone boson
or $4/3$ for a massive $U(1)$ gauge boson.
Finally, if a hot relic decouples earlier than the active neutrinos, it does
not receive any corrections from $e^+e^-$ annihilation. Since this is the case of interest, the pre-factor is smaller by $3/\Neff^\text{sm}$.
There is no factor of $1/3$, because the sterile neutrino possesses three times the mass of one
degenerate neutrino in~\eqref{rhonu}.

The current evidence for $\mnuseff > 0$ is mainly driven by galaxy cluster data~\cite{Wyman:2013lza,Hamann:2013iba} and supported by galaxy~\cite{Hamann:2013iba,Battye:2013xqa} as well as CMB lensing data~\cite{Battye:2013xqa}. Even before the Planck cluster count became available, cluster data has pushed evidence for $\mnuseff > 0$~\cite{Hou:2012xq,Burenin:2013wg}.
Corresponding HDM fractions $f$ are large enough to suppress the amplitude of matter fluctuations measured in  $\s_8$ (= root-mean-square fluctuation in total matter in $8 \, h^{-1} \Mpc$ spheres at $z=0$, computed in linear theory) down to values inferred from these ``local'' observations.
At the same time they appear small enough not to spoil the CMB fit, see also the next paragraph.

3) The average momentum of a sterile neutrino population is lowered (or increased) compared to
the active neutrinos by their temperature ratio $T_{\nu_s}/T_\nu$.
Comparing with~\eqref{Tnrnu1} and~\eqref{Tnrnu2}, we find
\begin{equation}
\label{Tnrnus}
 \Tnr_{\nu_s} = \frac{180 \z(3)}{7 \pi^4} \fb{11}{4}{\frac{1}{3}} \frac{T_{\nu}}{T_{\nu_s}} \mnus 
 \simeq 0.445 \,  \frac{\mnuseff}{\D\Neff}
\end{equation}
Now, there is a factor of three for
the reason given two paragraphs above.
For completeness, we insert into~\eqref{knr} obtaining
\begin{equation}
 k^\text{nr}_{\nus} \simeq 0.0178 \, \O_\text{m}^\frac{1}{2} \fb{\mnuseff}{\text{eV}}{\frac{1}{2}} \D\Neff^{-\frac{1}{2}} h \Mpc^{-1}  
\end{equation}
and
\begin{equation}
 \lfs_{\nus} \simeq 8.1 \, \O_\text{m}^\frac{1}{2} \D\Neff \frac{\text{eV}}{\mnuseff}  h^{-1} \Mpc .
\end{equation}

In Fig.~\ref{fig:posterior} we re-plotted the countours found in~\cite{Hamann:2013iba}.
They lie roughly around the line $\Tnr_{\nus} = T_\g^\text{dpl}$ with an upper bound on $\D\Neff$.
If a population becomes non-relativistic before photon decoupling, $\Tnr > T_\g^\text{dpl}$,
it affects the CMB directly, see e.g.~\cite{Dodelson:1995es}.
The non-observation of such an impact constrains the size of $\mnuseff$, in particular, for large $\D\Neff$. For smaller $\D\Neff$ the impact on the CMB decreases accordingly which allows for somewhat larger $\Tnr_{\nus}$. However, the observations appear to be matched for the obtained best-fits.
Even though $\Tnr_\nu$ and $\Tnr_{\nus}$ differ, there is neither preference in the data for $\sumnu >0$ with an additional source for ``dark radiation'' implying $\D\Neff>0$ nor for \nus HDM.
As only difference to HDM, dark radiation is still relativistic today. 
For such cosmologies parameters are obtained from our work in the limit $m_\text{hdm}\text{, }m_1\text{, }m_2 \rightarrow 0$.
\section{Perfect(?) Mimicry}
\label{sec:mimicry} 
We consider the two-body decay of a non-relativistic particle (matter or dust in the cosmological sense).
We distinguish between ``early'' cosmological decays occurring before the onset of the BBN era, so at a time $\t<0.05 \seconds$, and ``late'' decays occurring during or after the BBN era with $\t>0.05 \seconds$.
A decay during the CMB era deserves a dedicated study investigating
its impact on the CMB power spectrum. Therefore, we restrict ourselves to decay
times $\t \lesssim 5.2 \times 10^{10} \seconds$. This is before the first
observable  modes of the CMB enter the horizon~\cite{Fischler:2010xz}. 

In principle, a decay as drawn schematically in Fig.~\ref{fig:dof} might have to be described 
by an arbitrarily large number of parameters: mass $m$ and yield $Y$ of the decaying particle (mother) and the masses
of the decay products (daughters) $m_{1}^i$, $m_{2}^i$ in each existing decay mode $i$ that sum up to the 
total width  $\G= \sum_i \G_\text{vis}^i + \sum_i \G_\text{dark}^i= \t^{-1}$, where $\G_\text{vis}^i$ and $\G_\text{dark}^i$ denote partial decay widths into particles with and without electromagnetic interactions, respectively. 
Thus one might naively expect that there is an arbitrarily large number of possibilities for
the daughters to show up in the outlined physical effects.
In the following, we argue that, as far as the cosmological observables are concerned,
the decay can be described by a decisively smaller number of parameters

{\bf Early decay --}
The case of interest is the decay of a mother that dominates the energy density of the Universe at her decay (the opposite is covered in the next subsection) and produces HDM in its decay  that does not thermalise with the SM bath.
The majority of the energy stored in the mother must be transferred to visible particles that unavoidably thermalise. The information how the mother decayed into visible particles --contained in all $\G_\text{vis}^i$-- is irrelevant. Instead, the following cosmology will depend on branching ratios $B_i$ into hot dark matter $B_\text{hdm}$ and SM particles $B_\text{vis}$ only as they fix how the mother's energy is distributed. We just argued $B_\text{vis}\simeq 1 \gg B_\text{hdm}$. If one allows for (additional) decay modes to produce the observed CDM, the corresponding branching ratio were restricted to be much smaller than the one into HDM, because the CDM number density needs to be much smaller. So $B_\text{vis} = 1- B_\text{hdm}$ is either exact, if there is no additional dark decay mode, or holds to a sufficient approximation.

Concerning the kinematics, in the case of mass-degenerate daughters, $m_1 =m_2$, we define $x_2 \equiv m_2/m = m_1/m$ as measure of the mass hierarchy between mother and daughters.
It might well be that the well-motivated
case of daughters with similar masses, $m_1 \lesssim m_2$, is indistinguishable in cosmological observations from the case of mass-degenerate daughters. This is a subtle case that we leave for future work.
If the daughters possess a large mass hierarchy, $m_1 \ll m_2$,
the interesting case would be that there is a heavier daughter forming (observable)
HDM and an effectively massless one not doing so and thus forming dark radiation.
In that case the number of daughters forming HDM $g_\text{hdm}=1$, in contrast to the mass-degenerate case with $g_\text{hdm}=2$.
 We define $\d \equiv ( m - m_2)/m_2$ as measure of the mass hierarchy between mother
and heavier daughter.
Altogether, we can describe the kinematics approximately using only one hierarchy measure, $x_2$ or $\d$, depending on the case. 

The mother's energy density together with her lifetime determine the temperature of the thermal bath after her decay $\Trh$ as\footnote{
We define this temperature as the temperature of the standard thermal bath at the time of decay $T(\t)|_\text{rad-dom}$ calculated in radiation domination. This temperature is known as ``reheating'' temperature, therefore, the subscript. However, the Universe is not re-heated as at the end of inflation. It just cools more slowly during the decay period. From this point of view, the notation  used in the next subsection, $\Tdec$, might be seen as appropriate for this case, too. Anyway, to prevent confusion we distinguish the two cases explicitly also in the notation. 
}
\begin{equation}
 \label{rhomotherinearlydec}
\r|_\text{dec}= m n|_\text{dec}= \mu^{-1} \frac{\pi^2}{30}  \gast^\text{rh}  B_\text{vis}^{-1} \Trh^4 \, ,
\end{equation}
where the correction factor $\m = \m_\text{dom}\simeq 0.877$ considers the exponential decay law in the expanding Universe. We determined it by solving corresponding Boltzmann equations numerically in the limit of strong dominance, $\rho \gg \r_\text{rad}$, cf.~\cite{Scherrer:1987rr}.
Superscripts at $\gast$ --or, if the entropy density $s$ is considered, $\gasts$-- always indicated the temperature at which the functions are evaluated.

From an, in principle, arbitrarily large number of degrees of freedom we are down to four: $\Trh$, $B_\text{vis}$ (or $B_\text{hdm}$), $x_2$ or $\d$ and potentially $m$.
For an early decay the cosmological parameters under consideration are given by:

1) The relativistic HDM energy density $\r_\text{hdm}|_\text{rel} = n_\text{hdm} \langle p \rangle \simeq n_\text{hdm} (\m/2) m$,  where it has been exploited that the daughters must be much lighter than the mother due to structure formation constraints~\cite{Hasenkamp:2012ii}. With $n_\text{hdm} = g_\text{hdm} n$ it can be written as  $\r_\text{hdm}|_\text{rel} = \m_\text{dom} \r|_\text{dec} B_\text{hdm} (g_\text{hdm}/2) (T/\Trh)^4 (\gasts/\gasts^\text{rh})^{4/3}$.
Inserting~\eqref{rhomotherinearlydec} we see that 
\begin{equation}
 \label{DNeffearlydec}
\D\Neff = \frac{\r_\text{hdm}|_\text{rel}}{\r_{1\n}} \simeq 8.67 \, \frac{B_\text{hdm}}{B_\text{vis}} \fb{\gastst}{\gast^\text{rh}}{\frac{1}{3}} \, .
\end{equation}

2) Its today's (non-relativistic) energy density $\r_\text{hdm}^0 = n^0_\text{hdm} m_\text{hdm}$  can be written as $\r^0_\text{hdm}= \r|_\text{dec} x_2 g_\text{hdm} B_\text{hdm} (T_0/\Trh)^3 (\gastst/\gasts^\text{rh})$. Inserting~\eqref{rhomotherinearlydec} we see that it reads
\begin{equation}
 \label{Ohdmearlydec}
\O^0_\text{ed} h^2 = \frac{\r_\text{hdm}^0 h^2}{\rhoc} \simeq 4.15 \, \mu_\text{dom}^{-1} \frac{x_2}{10^{-8}} \frac{B_\text{hdm}}{B_\text{vis}}  \frac{\Trh}{\text{GeV}}\, .
\end{equation}

3) The temperature of the Universe when the population of free-streaming daughters becomes non-relativistic is found as
\begin{eqnarray}
 \Tnr_\text{ed} &=& \Trh \frac{2}{\mu} \frac{\d+1}{(\d+1)^2 -1} \fb{\gasts^\text{rh}}{\gastst}{\frac{1}{3}} \nonumber \\
\text{ or } \, &=& \Trh \frac{2}{\mu} (x_2^{-2} -4)^{-\frac{1}{2}} \fb{\gasts^\text{rh}}{\gastst}{\frac{1}{3}} ,
\label{Tnrdecay}
\end{eqnarray}
see (8) and (50) of~\cite{Hasenkamp:2012ii}, while $\m = \m_\text{dom}$, here. %
In all cases $\Tnr_\text{dec}< \Teq$ and the minimal comoving free-streaming wavenumber of the daughters is set by~\eqref{knr} with
$\Tnr$ given by~\eqref{Tnrdecay}.

By inspection of~\eqref{DNeffearlydec} and~\eqref{Ohdmearlydec} we see that the mother's mass does not enter independently. Furthermore, the observables depend on the ratio of the branching ratios $B_\text{hdm}/B_\text{vis}$ only. In other words, they depend on the amount of HDM relative to the amount of visible matter at the decay. This reduces the number of degrees of freedom in the description of the early decay to three: $\Trh$, $B_\text{hdm}/B_\text{vis}$ and a hierarchy measure, either $\d$ or $x$, depending on the case under investigation.

A dominating mother produces significant entropy that dilutes relic densities during her decay. The corresponding dilution factor $\D$ can be given as~\cite{Hasenkamp:2010if}
\begin{equation}
\D = \frac{\langle \gast^\frac{1}{3} \rangle^\frac{3}{4}}{(\gast^\text{rh})^\frac{1}{4}} \frac{mY}{\Trh} \, ,
\end{equation}
where the angle brackets indicate the appropriately-averaged value of $\gast$ over the decay interval. We see that $\D$ is independent of the HDM observables. The dilution is an \textit{additional, independent} physical effect. After discovering the primordial gravitational wave background in the CMB, we \textit{know} that this impact of the decay will be ultimately tested by a local detection of this background radiation~\cite{Durrer:2011bi} that would rule out any such early period of matter domination .

{\it \bf Late decay --}
For $\t > 0.05 \seconds$ current constraints on $\D\Neff$ forbid the particle to dominate the energy
density of the Universe prior its decay, see Sec.~2.1 of~\cite{Hasenkamp:2012ii}.\footnote{
Of course, the following considerations hold for any earlier decay of a non-relativistic particle that does not dominate at its decay. There is just no observational reason that would forbid its domination.
}
For such late decays, if the mother is sufficiently heavier
 than the proton, the branching ratio into any visible particles is constrained
from BBN and CMB observations to be much smaller 
than one, $\sum_i \G_\text{vis}^i \ll  \sum_i \G_\text{dark}^i$, see Sec.~3 of~\cite{Hasenkamp:2012ii}.
This holds even if photons and electrons are emitted at the end of a decay chain only, cp.~\cite{Cline:2013fm}.
If one allows for (additional) decay modes to produce the observed dark matter,
the corresponding branching ratio were restricted to be much smaller than the one into
hot dark matter, $\sum_i \G_\text{dark}^i \simeq \G_\text{hdm}$, see Sec.~4.2 of~\cite{Hasenkamp:2012ii}.
In other words, the HDM branching ratio is to a very good approximation one
and the lifetime of the mother is given by her HDM decay width.
This implies a decisive reduction of parameters. From an, in principle, arbitrarily large number
we are down to four: mass $m$, yield $Y$ and lifetime $\t$ of the mother and one of the two hierarchy measures, $\d$ or $x_2$.

For the late decay following~\cite{Hasenkamp:2012ii}
the cosmological parameters under consideration are given by:

1) If $\Tdec= T(\t)$ denotes the
temperature of the Universe at decay\footnote{
The Universe is radiation dominated in the time window under consideration.
},
\begin{equation}
\label{DNeffdecay}
 \D\Neff = 10.25 \, \frac{m Y}{\Tdec} \fb{\gastst}{\gastsd}{\frac{1}{3}} \frac{(\d+1)^2-1}{(\d+1)^2} \, .
\end{equation}
Again taking into account the huge mass hierarchy, $\d\gg 1$, required from structure formation constraints,
the last fraction is also in this case to a good approximation one. This holds analogously, if the daughters are mass degenerate, see~(20) of~\cite{Hasenkamp:2012ii}.

2) The non-relativistic energy density of the daughters today is given by, cf.~(45) and~(55) of~\cite{Hasenkamp:2012ii},
\begin{align}
\label{Odecay}
\O_\text{ld}^0 &= g_\text{hdm} \frac{mY}{\d +1} \frac{\gastsd}{\gastst} \frac{s_0}{\rhoc}  \nonumber \\
& \Leftrightarrow \O_\text{ld}^0 h^2 = \frac{2.76 \times 10^8}{\d+1} \frac{mY}{\text{GeV}} g_\text{hdm} \frac{\gastsd}{\gastst}\, .
\end{align}
For a decay into mass-degenerate daughters we should replace $1/(\d +1) \rightarrow  x_2 $.
Note that the yield $Y$ is the one in~\eqref{DNeffdecay}. So it is to be evaluated at the mother's decay.
nto two identical particles we should replace $1/(\d +1) \rightarrow 2 x_2 = 2 m_2 /m$.
Note that by construction the same decay
leads to both, $\D\Neff>0$ and $\mnuseff>0$. We discuss the implications of a second
decay mode into dark radiation or HDM below.

3) The temperature of the Universe when the population of free-streaming daughters becomes non-relativistic is given by~\eqref{Tnrdecay} with $\m = \sqrt{\pi}/2$.

We inspect~\eqref{DNeffdecay} and~\eqref{Odecay} to complete our counting of degrees of freedom in the description of a late cosmological particle decay, for the moment. We see that both observables
depend on the product $mY$ only and not on $m$ or $Y$ independently. They depend on the energy
density at decay and not on how the energy density is built from a large mass or number density, as for an early decay.
Thus observations are not sensitive to this degeneracy in determining the energy density of the
mother.
So instead of an arbitrarily large number of degrees of freedom 
we identify three: the lifetime $\t$, the energy density of the decaying particle $\r(=mYs)$
and a hierarchy measure, either $\d$ or $x$, depending on the case under investigation.

Even though the descriptions for the early and late decay are qualitatively different, 
 we find in both cases three degrees of freedom that are possibly affecting the outlined three main physical effects of a 
free-streaming population. One might compare this to \nus HDM with
 two parameters, i.e., $\Tnus$ and $\mnus$. 
In contrast to \nus HDM, no (possibly unnatural) small mass scale needs to be introduced.

\subsection{Translating observations}
\begin{figure}
\includegraphics[width=\columnwidth]{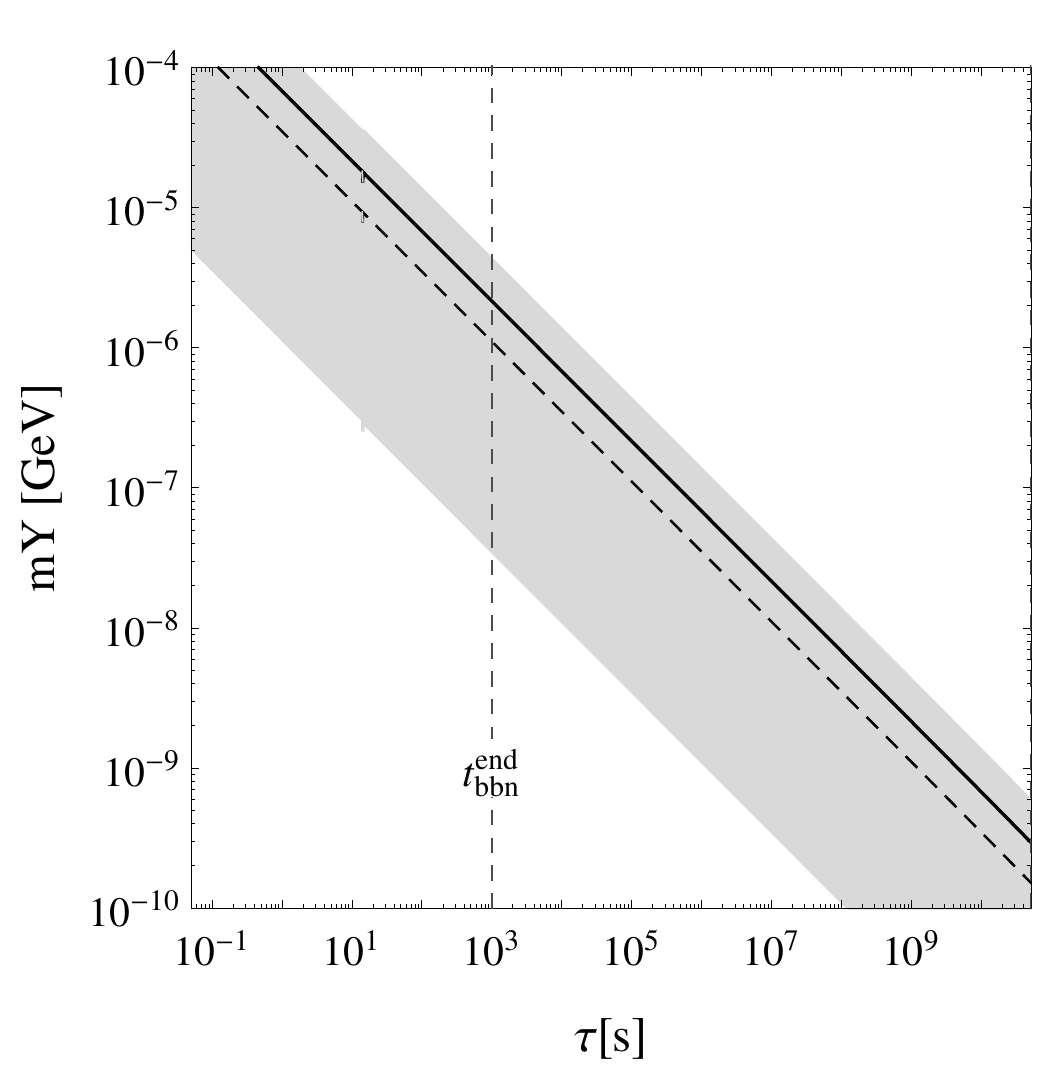} 
\label{fig:DNeffdecay}
\caption[{$mY(\t)$ fixed by $\Tnus$}]{ Determination of $mY(\t)$ depending on $\Tnus$ exploiting~\eqref{DNeffdecay}.
A higher $\Tnus$ is mimicked by a correspondingly larger energy density of the decaying particle at decay. 
The solid line corresponds to $\Neff = 0.61 \pm 0.60$ with its 2-$\s$ range as grey band.
The dashed line marks Planck's CMB only mean $\D\Neff=0.29$~\cite{Ade:2013zuv}.
}
\end{figure}
\begin{figure}%
\centering
\includegraphics[width=\columnwidth]{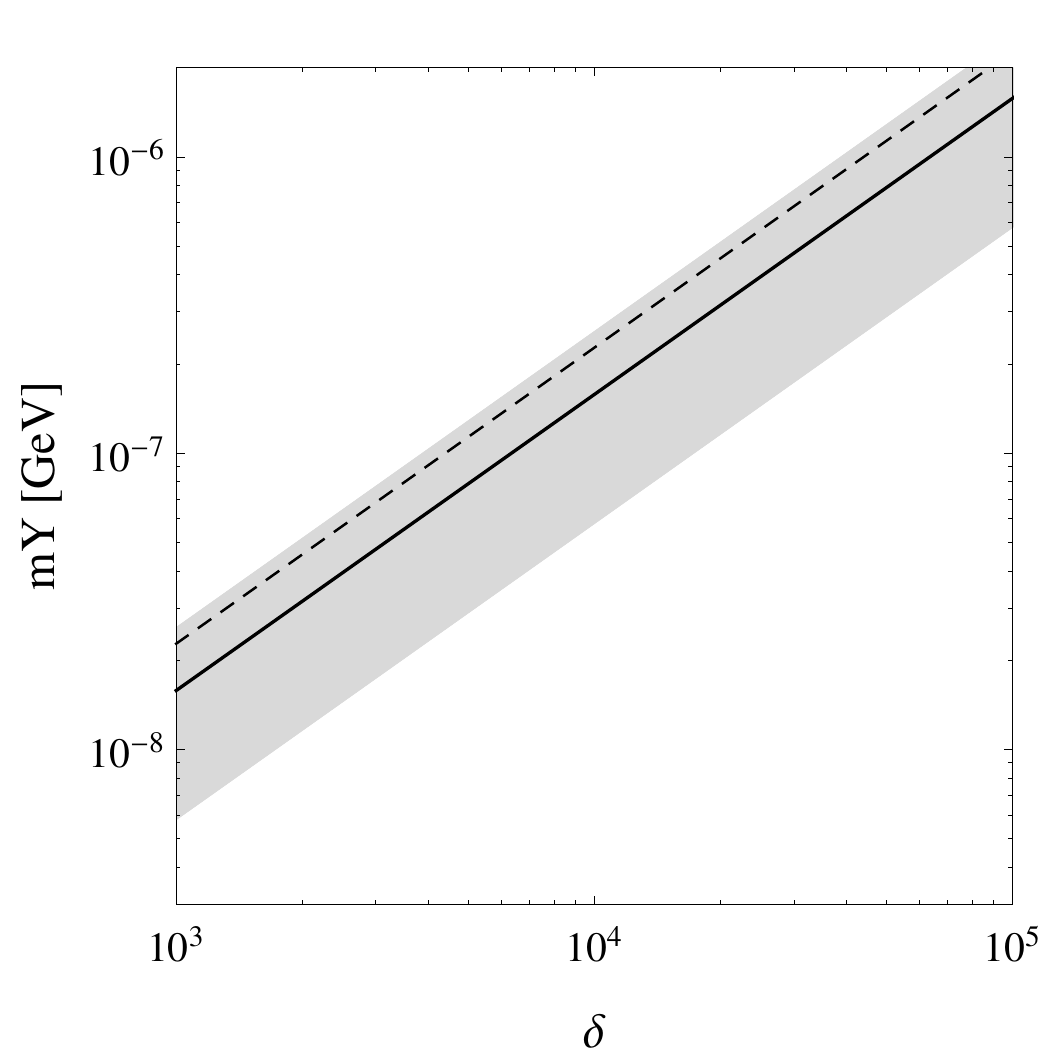} 
\label{fig:Odecay}
\caption{Determination of $mY(\d)$ depending only on $\mnuseff$.
A larger $\mnuseff$ is mimicked by a correspondingly smaller mass hierarchy.
The solid line corresponds to $\mnuseff/\text{eV} = 0.41 \pm 0.26$ with its 2-$\s$ range as grey band.
The dashed line marks Planck's CMB-only upper bound $\mnuseff < 0.59 \eV$ (95\%)~\cite{Ade:2013zuv}.
We have chosen $g_\text{hdm}=1$ for the figure. If the decay is into mass-degenerate daughters, the horizontal axis should carry $x_2^{-1}$ and all lines shift down by $1/2$ as $g_\text{hdm}=2$.
}
\end{figure}
\begin{figure}
\includegraphics[width=\columnwidth]{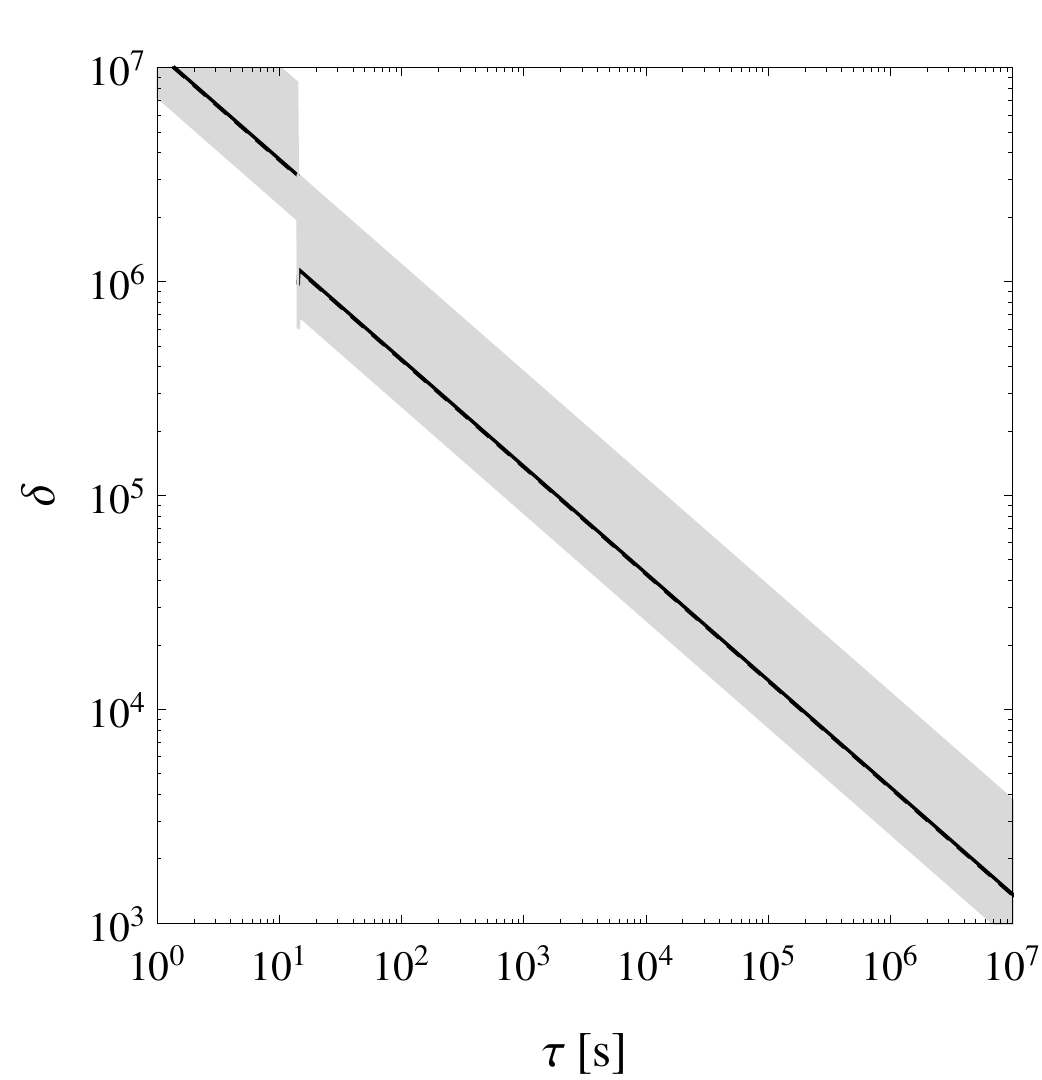} 
\label{fig:Odecay2}
\caption{Determination of $\d(\t)$ depending only on $\mnuseff$ 
exploiting~\eqref{DNeffdecay} and~\eqref{Odecay} with fixed $\Tnus$.
For a fixed $\mnuseff$ the later the decay the smaller the mass hierarchy might be.
The solid line corresponds to $\mnuseff/\text{eV} = 0.41 \pm 0.26$ with its 2-$\s$ range as grey band, while $\D\Neff=0.61$ is kept fix.
The jump around $t\sim 10 \seconds$ corresponds to the time of $e^+e^-$ annihilation.
Planck's CMB only upper bound $\mnuseff < 0.59 \eV$ and $\D\Neff < 0.86$ (both 95\%)~\cite{Ade:2013zuv} is barely visible right on the solid line.
We have chosen $g_\text{hdm}=1$ for the figure. If the decay is into identical particles, the vertical axis should carry $x_2^{-1}$ and all lines shift up by $1/2$ as $g_\text{hdm}=2$.
}
\end{figure}

In the following we show actually that dpHDM and \nus HDM are in general indistinguishable in CMB and large-scale structure observations using $\Neff$ and $\mnuseff$ only. 
First, we provide the prescription how to obtain 
(for any ``observed'' sterile neutrino parameter)
 the corresponding
 parameters describing the cosmological particle decay
with exactly the same signal.
In other words, parameters determined assuming a sterile neutrino cosmology
can be translated back and forth into parameters describing a cosmology with
HDM originating from a cosmological particle decay.

{\bf Early decay --}
Comparing~\eqref{DNeffnus} and~\eqref{DNeffearlydec} we see trivially that a higher sterile neutrino temperature is mimicked by a larger $B_\text{hdm}/B_\text{vis}$ (for fix $\gast^\text{rh}$).
Furthermore, equating~\eqref{Onus} and~\eqref{Ohdmearlydec} we find how the effective \nus HDM mass is mimicked,
\begin{equation}
\label{mnuseffearlydec}
 \mnuseff = \mu_\text{dom}^{-1} \frac{11 \pi^4 \gastst}{90 \zeta(3)} \frac{B_\text{hdm}}{B_\text{vis}} x_2 \Trh \, .
\end{equation}
So the prescription is: i) Fix $B_\text{hdm}/B_\text{vis}$ to obtain the desired $\D\Neff$. ii) Determine $x_2 \Trh$, while $\D\Neff$ is kept fix. At this point keep in mind that the description is valid for $\Trh > \MeV$ only. 
For illustration, we translate~\eqref{hdmsignal} into 
\begin{equation}
\label{sigDNeffearlydec}
 \frac{B_\text{hdm}}{B_\text{vis}} \left( \gast^\text{rh}\right)^{-\frac{1}{3}} = 0.045 \pm 0.022 
 \text{ } (\text{``1-}\s\text{''} \text{; } \t < 0.05 \seconds)
\end{equation}
and
\begin{align}
\label{sigmnuseffearlydec}
 \frac{B_\text{hdm}}{B_\text{vis}} \frac{x_2}{10^{-9}} \frac{\Trh}{\text{GeV}} =  (9.2 \pm 2.9) & \times 10^{-3} \nonumber \\ 
& (\text{``1-}\s \text{''; } \t < 0.05 \seconds) 
\end{align}
for a scenario with mass-degenerate daughters.
If the daughters possess a large mass hierarchy, such that $g_\text{hdm}=1$, 
a) nothing changes for $\D\Neff$ and 
b) $x_2 \rightarrow 1/(\d+1)$ in~\eqref{sigmnuseffearlydec} and its right-hand-side is to be divided by 2.
However, in this case there is a shift in $\Tnr$ as shown below and thus a difference in the main physical effects.

We would like to point out that~\eqref{sigDNeffearlydec} and~\eqref{sigmnuseffearlydec} (later also~\eqref{sigDNeff} and~\eqref{sigmnuseff}) are by way of illustration only. We simply ``translated'' parameters, while the confidence intervals are not determined by a likelihood ratio, but by the area below the posterior function that is not invariant under non-linear parameter transformations. 
In Sec.~\ref{sec:breakdown} we identify (two, independent) parameters that should be determined in a Markov Chain Monte Carlo likelihood analysis with approriate priors. 

{\bf Late decay --}
How to mimic a sterile neutrino in $\Neff$ is shown in Fig.~\ref{fig:DNeffdecay}.
The horizontal axis spans the full range of considered lifetimes.
We take the temperature dependence of $\gasts$ fully into account.
Around $\t\sim10 \seconds$, when $e^+e^-$ annihilate away, $\gasts$ decreases. The curves stay straight as for the same reason the temperature of the Universe increases, both effects cancelling each other up to the difference between $\gastt\simeq3.384$ and $\gastst\simeq3.938$. It is roughly $(\gastst/\gastt)^{1/3} \simeq 1.05$.
The time when BBN ends $t_\text{bbn}^\text{end}$ is highlighted.
Decays before this time increase the effective number of neutrino species inferred from
BBN $\D\Neff|_\text{bbn}$ partially as quantified in~\cite{Menestrina:2011mz}.
Comparing~\eqref{DNeffnus} and~\eqref{DNeffdecay} we see that a higher $\Tnus$
is mimicked by a larger energy density of the mother at decay.
This has been shown in~\cite{Hooper:2011aj} solving the corresponding Boltzmann equations numerically.
The values chosen for $\D\Neff$, which are in one-to-one correspondence with $\Tnus$ via~\eqref{DNeffnus},
 are given in the figure caption.
We can see how a narrow range of $\Tnus$ maps onto a narrow range of $mY(\t)$ forming a strip in the $mY$-$\t$ plane.

After having fixed $mY/\Tdec$ we can use Fig.~\ref{fig:Odecay}, which is obtained by setting~\eqref{Onus} equal to~\eqref{Odecay}, to determine the correct $\d(mY)$ 
mimicking the effective HDM mass 
\begin{align}
 &\mnuseff \simeq 26.0 \, g_\text{hdm} \frac{mY}{\d+1} \frac{\gastsd}{\gastst} \nonumber \\
&\text{ or } \simeq 52.0 \, x_2 mY  \frac{\gastsd}{\gastst} \, . 
\end{align}
Alternatively, we can use~\eqref{DNeffdecay} to re-write~\eqref{Odecay}, which equals~\eqref{Onus}, obtaining
\begin{equation}
\label{deltaofobs}
 \d \simeq 5 \times 10^3 \,  \frac{\Tdec}{\text{keV}} \D\Neff \frac{\text{eV}}{\mnuseff}
\fb{\gastsd}{\gastst}{\frac{4}{3}} \frac{g_\text{hdm}}{2} -1 
\end{equation}
which allows to provide Fig.~\ref{fig:Odecay2}.
For a given lifetime we can thus read off the corresponding $mY$ and $\d$ to
mimic any ``observed'' sterile neutrino parameters.

Altogether, we found the following prescription:
i) Determine $mY(\t)$ using~\eqref{DNeffdecay} such that the desired $\D\Neff$ is obtained. 
ii) Determine $\d(\t)$ using~\eqref{deltaofobs} such that the desired $\mnuseff$ is obtained while $\D\Neff$ is kept fixed.
This makes dpHDM indistinguishable from \nus HDM
in $\Neff$ and $\mnuseff$.
For illustration, we translate~\eqref{hdmsignal} into 
\begin{align}
\label{sigDNeff}
 \frac{mY}{\Tdec}\fb{\gastst}{\gastsd}{\frac{1}{3}} &= 0.060 \pm 0.029 \nonumber \\
& (\text{``1-}\s \text{''; } 0.05 \seconds <\t <5.2 \times 10^{10} \seconds)
\end{align}
and
\begin{align}
\label{sigmnuseff}
 x_2 \, mY \frac{\gastsd}{\gastst} &= (7.9 \pm 2.5)\times 10^{-3} \eV  \nonumber \\
&(\text{``1-}\s \text{``; } 0.05 \seconds <\t <5.2 \times 10^{10} \seconds)
\end{align}
for a scenario with mass-degenerate daughters.
If the daughters possess a large mass hierarchy, such that $g_\text{hdm}=1$, proceed as discussed below~\eqref{sigmnuseffearlydec}.

It is important that --in contrast to the early decay case-- the assumption of BBN consistency is wrong, because for a late cosmological decay $\Neff|_\text{bbn} < \Neff|_\text{cmb}$. The possibility of late cosmological decays motivates determinations of $\Neff|_\text{cmb}$, where the primordial abundance of light elements is constrained from direct observations only, so \textit{really} independent of cosmology at earlier times, and fixed $\Neff|_\text{bbn}=\Neff^\text{sm}$. For earlier related work with a different approach in a different scenario see~\cite{GonzalezGarcia:2012yq}.

{\bf A degeneracy --}
For the early decay we can exploit~\eqref{DNeffearlydec},~\eqref{Tnrdecay} and~\eqref{mnuseffearlydec} to obtain
the following \textit{parameter degeneracy} relation
\begin{equation}
 \label{Tnrobs2}
\Tnr_\text{ed} = \frac{180 \zeta(3)}{7 \pi^4} \fb{11}{4}{\frac{1}{3}} \frac{\mnuseff}{\D\Neff} \frac{2}{g_\text{hdm}} \, .
\end{equation} 
Analogously, if we use~\eqref{DNeffdecay} and~\eqref{Odecay} eliminating $mY$, we can single out 
$\Tdec (\d+1)/ ((\d+1)^2 -1)$. Inserting the resulting expression into~\eqref{Tnrdecay} we obtain
\begin{equation}
  \label{Tnrobs}
\Tnr_\text{ld} \simeq 0.445 \frac{\mnuseff}{\D\Neff} \frac{\gastst}{\gastsd} \frac{2}{g_\text{hdm}} \, .
\end{equation}
We see that the main physical effects are described by only two independent parameters.
In other words, there is a degeneracy among the three physical parameters.
The same degeneracy has been identified for the thermal relic case in~\cite{Acero:2008rh}.\footnote{
It has been used to show that a Dodelson-Widrow model shares the same ''observable`` parameters
as a thermal sterile neutrino model with adjusted mass and temperature. 
}
Indeed, we find exactly the same degeneracy, i.e. the same numerical value, if we
consider $g_\text{hdm}=2$ and -for the late decay- $\gastsd = \gastst$, which is fulfilled after $e^+e^-$ annihilation, so in a large part of the parameter space.
Viewed as a constraint equation the degeneracy in physical parameters reduces
the number of independent degrees of freedom describing dpHDM from three to two. 
Considering $\Neff$ and $\mnuseff$ only, which can describe the
 main physical effects, 
an early cosmological particle decay, as well as a decay after $e^+e^-$ annihilation, as origin of
HDM is indistinguishable from \nus HDM, if the daughters are mass degenerate.
The mimicry is ''perfect`` in the sense that there is no new signature
in the main physical effects as those contained in \nus HDM parameters ($\Tnus$ or $\D\Neff$ and $\mnus$ or $\mnuseff$).

{\it \bf More complex scenarios}
If the daughters do not possess a large
mass hierarchy, $m_1\lesssim m_2$, their impact on observables is similar to the case
of two sterile neutrinos with equal temperature but different mass.
Moreover, considering two sterile neutrino populations with sufficiently different temperatures, one neutrino might not account for 
 $\O_\text{hdm}$ as being still relativistic today and thus forming dark radiation. This situation would correspond to
two dark decay modes in the cosmological particle decay, one for dark radiation
and one for HDM. All we would like to point out here is that these scenarios
are more complex also in the case of a thermal relic. Consequently, they are not discussed in this work and left for future work.

The arguably most interesting case is found if the heavier daughter forms HDM and the lighter one dark radiation. Then the temperature of the universe, when the
HDM becomes non-relativistic is a factor of two larger as $g_\text{hdm}=1$ in~\eqref{Tnrobs2} and~\eqref{Tnrobs}, respectively.
Translated to sterile neutrinos, this situation is represented by two sterile neutrino populations with the same temperature, while one is massive and the other one is massless.
Interestingly, current (non-zero) mean values for a HDM contribution correspond
to $\Tnr$s around $T_\g^\text{dpl}$, see Fig.~\ref{fig:posterior}.
If $\Tnr$ becomes significantly larger than $T_\g^\text{dpl}$ due this factor of two,
this effect could be used to distinguish this case, because the impact on the CMB
becomes qualitatively different for $\Tnr >T_\g^\text{dpl}$~\cite{Dodelson:1995es}.
As mentioned already, the non-observation of such an impact constrains the size of $\mnuseff$.
While we do not attempt quantitative statements, here, it seems that the increase in $\mnuseff$ would be in stronger tension with the CMB data for $g_\text{hdm}=1$. This could disfavour this case compared to the case of mass-degenerate daughters.
On the other hand, if a late decay occurs before $e^+e^-$ annihilation, the daughters become non-relativistic at a time $\gastsd/\gastst \simeq 2.73$ later which reduces the constraining power from the direct impact on the CMB.

For lower $\Tnr$ it has been argued in~\cite{Hannestad:2005bt} that an optimal LSST-type wide field survey might provide the ability to distinguish
between thermal, fermionic HDM (Majorana fermion) and thermal, bosonic HDM (scalar).
For fixed $\O_\text{hdm}$ the scalar becomes non-relativistic at a temperature,
which is a factor of $3/2$ larger than in the case of a fermion. This is simply
the increase in mass compensating the factor considering the scalar particle nature in~\eqref{Onus}.
We might conclude that the same observations are able
to differ between \nus HDM and dpHDM where the heavier daughter forms HDM and the lighter one dark radiation. However, on the observed non-linear scales N-body simulations appear to become necessary~\cite{Bird:2011rb} and HDM infall might become an observable effect with the increased precision~\cite{Hannestad:2005bt}.
The exploration of this interesting opportunity to probe the origin of HDM is beyond the scope of this work.
In future work, all discussed scenarios can be implemented in CAMB\footnote{
http://camb.info/
} or CLASS\footnote{
http://class-code.net/
} easily, because they can be represented by thermal fermion populations as just described and summarised in Tab.~\ref{tab:summary}.
\subsection{Breakdown in the Third}
\label{sec:breakdown}
\begin{figure}
\includegraphics[width=\columnwidth]{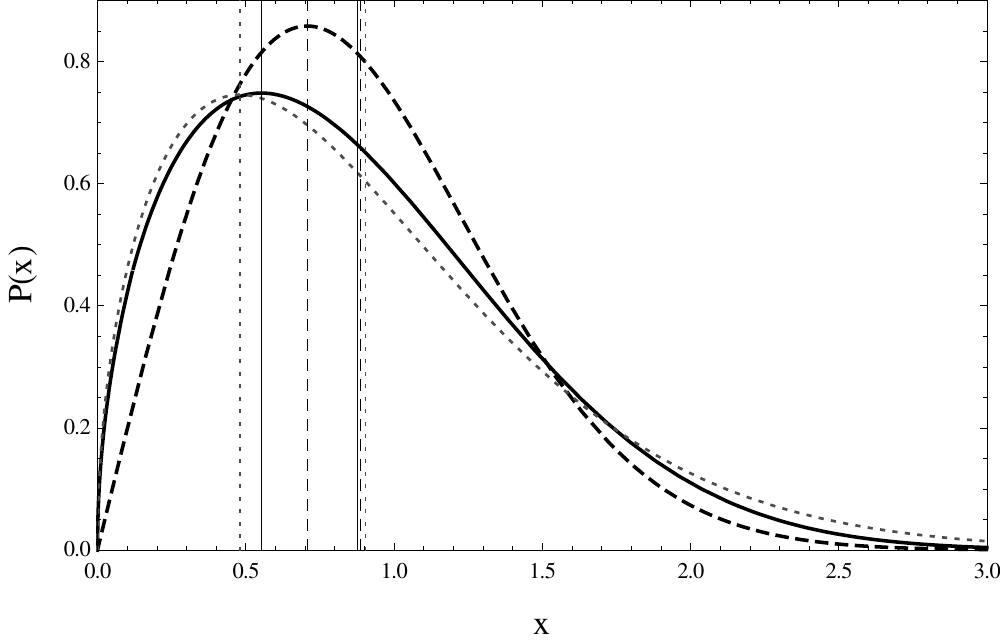} 
\caption{Normalised, time-invariant probability distribution $P(x=p/p(\t,t))$ for finding a relativistic particle from two-body decay (decaying particle at rest) within the infinitesimal momentum interval $[p,p+\d p]$. In the early decay case the mother may dominate (black, solid) the Universe. For a late decay the Universe is radiation dominated (black, dashed). The Universe becomes matter-dominated (grey, dotted) at very late times only. 
Highlighted with corresponding lines are the maxima and means.
}
\label{fig:Px_overplot}
\end{figure}
It is a simplification to reduce the physical impact of any population of massive free-streaming particles
to the effects listed in Sec.~\ref{sec:preliminaries}. The free-streaming
effect must depend on the details of the momentum distribution $f(p)$ of the population.
For illustration imagine a distribution with a sharp peak close to $p=0$. Such particles would
act as cold dark matter and thus should not be counted within the massive free-streaming component~\eqref{Onus}.
Furthermore, the average considered in~\eqref{defTnr} were between cold and hot particles, so that
the correct physical effects were not captured.

{\bf Observables and statistical moments --}
Distribution functions can be specified by their moments $Q^{(n)}$. 
This has been done for the Fermi-Dirac
(active neutrino) distribution $f_\text{fd}(y)=1/(e^y +1)$ with comoving momentum $y=pa$ in~\cite{Cuoco:2005qr}.
As their result can be adapted to the sterile neutrino case straightforwardly, we 
 sketch decisive steps in Appendix~\ref{appendix:nudistr}.
For a thermal population of sterile neutrinos we define
\begin{equation}
\label{Qnus}
 Q^{(n)}_{\nus} = \frac{1}{\pi^2} \fb{4}{11}{\frac{3+n}{3}} \fb{\Tnus}{T_\nu}{3+n} T^{3+n} 
\int y^{2+n} f_\text{fd}(y) dy \, .
\end{equation}
The (first two) cosmological parameters can be expressed by these moments as
\begin{equation}
\label{NeffbyQnus}
 \Neff = 3.046 + \frac{120}{7 \pi^2} \fb{11}{4}{\frac{4}{3}} T^{-4} Q^{(1)}_{\nus}
\end{equation}
and
\begin{align}
\O_{\nus}^0 &= \frac{\mnus}{\rhoc} \fb{T_0}{T}{3} Q_{\nus}^{(0)}  \Leftrightarrow \nonumber \\
  \O_{\nus}^0 h^2 & \simeq 0.160 \, \frac{\mnuseff}{\text{eV}} \D\Neff^\frac{3}{4} T^{-3} Q_{\nus}^{(0)} \, .
\end{align}
We see that $\Neff$ and $\O_{\nus}^0$ probe the first two moments of the momentum distribution
function.

If the mother does not dominate the Universe at her decay,
the momentum distribution function of relativistically emitted daughters (cf.\@ appendix of~\cite{Hasenkamp:2012ii} and~\cite{Scherrer:1987rr})
as function of comoving momentum $y$ reads
\begin{equation}
\label{decdistr}
 df(y)=  \nini c  p_\text{ini}^{-c} y^{c-1} e^{-(y/p_\text{ini})^c} dy \, ,
\end{equation}
where $n_\text{ini}$ denotes the number density of daughters at decay and $p_\text{ini}= m (1-(\d+1)^{-2})/2$
 or $=m (1 - 4 x_2^2)^{1/2}/2$ their initial momentum. For the considered lifetime range
it is well approximated by $p_\text{ini} \simeq m/2$.
In Fig.~\ref{fig:Px_overplot} we depicted the corresponding probability distribution function.

For the late decay
we use explicitly that the Universe is radiation dominated, so that $c\equiv 3(1+\o)/2 =2$ as the equation of state
of the Universe, $p = \o \r$, is given by $\o=1/3$ in that case. 
Nevertheless, our treatment could be applied to any expansion law with $c>0$.
By inspection of~\eqref{decdistr} we can see that the distribution function decays as $e^{-y}$ for large comoving momenta as the
neutrino distribution function considered in~\cite{Cuoco:2005qr}.
We define the moments as
\begin{equation}
\label{Qdec}
 Q_\text{dec}^{(n)} = \frac{\nini}{\Tdec^3} \fb{\pini}{\Tdec}{n} \fb{\gastst}{\gastsd}{\frac{n}{3}} T^{3+n} \int c y^{c+n-1} e^{-y^c} dy 
\end{equation}
Then the (first two) cosmological parameters expressed by these moments read
\begin{equation}
 \Neff = 3.046 + \frac{120}{7\pi^2}\fb{11}{4}{\frac{4}{3}} T^{-4} Q_\text{dec}^{(1)}
\end{equation}
and
\begin{align}
\label{OQdec}
\O_\text{dec}^0 &= \frac{g_\text{hdm}}{2} \frac{m_2}{\rhoc}  \fb{T_0}{T}{3} Q_\text{dec}^{(0)} \Leftrightarrow \nonumber \\
\O_\text{dec}^0 h^2 &\simeq 0.160 \, \frac{g_\text{hdm}}{2}  \frac{m_2}{\text{eV}} T^{-3} Q_\text{dec}^{(0)}
\end{align}
Note that $n_\text{ini}= 2 n$. We identify two dimensionless, independent parameters $\nini/\Tdec^3$ and $\pini/\Tdec$
in accordance with the previous discussion. 
In future work, these could be determined with a Markov Chain Monte Carlo likelihood analysis.
Our discussion from~\eqref{Qnus} to~\eqref{OQdec}
is an alternative way to demonstrate why the mimicry in $\Neff$ and $\mnuseff$ is ''perfect``.

In the case of a dominating mother, the simple analytic form~\eqref{decdistr} is invalid. Boltzmann equations need to be solved numerically, cf.~\cite{Scherrer:1987rr}. In Fig.~\ref{fig:Px_overplot} we depicted the resulting probability distribution in the limit of strong dominance, $\r/\r_\text{rad} \gg 1$, at decay.
Concerning the moments~\eqref{Qdec}, the factor $\gastst/\gastsd$ were absent and the integral could not be performed analytically. 

{\bf Root-mean-square of absolute momenta --}
The momentum distribution function of a thermal relic differs from the momentum distribution
function of relativistic decay products, cp.~\eqref{nudistr} and~\eqref{decdistr}. The two available degrees of 
freedom have been used to achieve perfect mimicry in the first two cosmological parameters, which have
been expressed by the first two moments of the distribution functions.
Therefore, the \textit{mimicry must break down} if the next (third) cosmological parameter, which is to be expressed
by the next higher moment of the distribution function, is taken into account. The number
of parameters to be mimicked then exceeds the number of degrees of freedom.

We identify the third cosmological parameter as the root-mean-square of absolute momenta $\langle p \rangle_\text{rms}$ --or, equivalently, the root-mean-square of absolute velocities today\footnote{
If clustering can be neglected, today's root-mean-square of absolute velocities $\langle v^0 \rangle_\text{rms} = \langle p^0 \rangle_\text{rms}/m$.
}--, which is given by the quadratic mean of the distribution of absolute momenta.
This quadratic mean is equal to the skewness of the momentum distribution keeping the directional information. 
In general, the skewness is the next higher moment following the quadratic mean. The next higher moment to the skewness is the kurtosis.
Giving preference to dimensionless parameters we define the \textit{normalised root-mean-square of absolute momenta}
\begin{equation}
 \label{defnormrmsmomentum}
\g^\text{rms} \equiv \frac{\langle p \rangle_\text{rms}}{\langle p_\nu \rangle_\text{rms}} 
\end{equation}
where $\langle p_\nu \rangle_\text{rms}$ is the root-mean-square of the absolute momentum of the active neutrinos~\eqref{rmsnumomentum}, so that $\g_\nu^\text{rms} = 1$ by definition.
The root-mean-squares of absolute momenta can be expressed by moments of the corresponding momentum distribution functions.
 This can be understood by reminding that the number density $n = Q^{(0)}$ and the relativistic energy density $\r =\langle p \rangle n = Q^{(1)}$, so that the mean absolute momentum $\langle p \rangle = Q^{(1)}/Q^{(0)}$. Analogously, $\langle p \rangle_\text{rms} = (Q^{(2)}/Q^{(0)})^{1/2}$.

For a thermal sterile neutrino population we find
\begin{equation}
 \label{nusprms}
\langle p_{\nu_s} \rangle_\text{rms}
= \fb{15 \zeta(5)}{\zeta(3)}{\frac{1}{2}} \fb{4}{11}{\frac{1}{3}} \frac{\Tnus}{T_\n} T \simeq 3.6 \, \Tnus \,,
\end{equation}
which implies with~\eqref{rmsnumomentum} a normalised root-mean-square of absolute momenta
\begin{equation}
\label{grmsnus}
 \g_{\nu_s}^\text{rms} = \frac{\Tnus}{T_\n} = \D\Neff^{\frac{1}{4}}
\end{equation}
that can be expressed by lower momenta~\eqref{NeffbyQnus}. This is to be expected, since $\g_{\nu_s}^\text{rms}$ cannot carry additional information as both neutrino species possess a Fermi-Dirac distribution.

The momentum distribution function of the sterile neutrinos in the popular Dodelson-Widrow (DW) scenario~\cite{Dodelson:1993je} (a.k.a.\@ ''non-resonant production scenario``) reads
\begin{equation}
 f_\text{dw}(y) = \frac{\chi}{e^y+1} = \chi f_\text{fd} \,.
\end{equation}
The sterile neutrinos share the same ''observable`` parameters ($\D\Neff$, $\O_\text{hdm}^0$, $T^\text{nr}$) as a thermal population with $\mnus^\text{th} = \chi^{1/4} \mnus^\text{dw}$ and $\Tnus^\text{th} =\chi^{1/4} T_\nu$.
For these two models it has been shown by a change of variable in the background and linear perturbation equations that the two models are strictly equivalent for cosmological observables~\cite{Colombi:1995ze}.
We confirm this result with the expansion of the distribution functions in moments. Compensating the scaling factor $\chi$, the two models share the same (Fermi-Dirac) distribution function. Higher moments can be expressed by lower ones and the distribution functions are identical. Therefore, they are indistinguishable for cosmological observations even if higher moments are taken into account. In other words, their \textit{mimicry is perfect to arbitrary order in moments}. 

This is \textit{qualitatively different for the decay products} of a cosmological particle decay mimicking a thermal sterile neutrino population.
For daughters emitted in a late decay we find
\begin{equation}
 \langle p_\text{dec} \rangle_\text{rms} =
\fb{Q_\text{dec}^{(2)}}{Q_\text{dec}^{(0)}}{\frac{1}{2}} 
= \frac{\pini}{\Tdec} T  \fb{\gastst}{\gastsd}{\frac{1}{3}} \, ,
\end{equation}
where we used that $\int c y^{c+n-1} e^{-y^c} dy =1 $ for $c=2$ in both cases, $n=0$ and $n=2$. We obtain a very simple, exact expression.
With~\eqref{rmsnumomentum} this implies
\begin{eqnarray}
 \g_\text{ld}^\text{rms} &=&  
\fb{\zeta (3)}{15 \zeta (5)}{\frac{1}{2}}
\fb{11}{4}{\frac{1}{3}}
\fb{\gastst}{\gastsd}{\frac{1}{3}}
\frac{\pini}{\Tdec} \nonumber \\
&\simeq& 0.389 \, \frac{\pini}{\Tdec} \fb{\gastst}{\gastsd}{\frac{1}{3}}  ,
\end{eqnarray}
which can be expressed as
\begin{equation}
\label{grmsdec}
\g_\text{ld}^\text{rms} \simeq 0.0105 \, \frac{\D\Neff}{\O_\text{ld}^0 h^2} \frac{g_\text{hdm}}{2}
\end{equation}
by lower moments. Again, this must be, since there are no further parameters. All higher momenta are fixed by the first two. 

\begin{figure}
\includegraphics[width=\columnwidth]{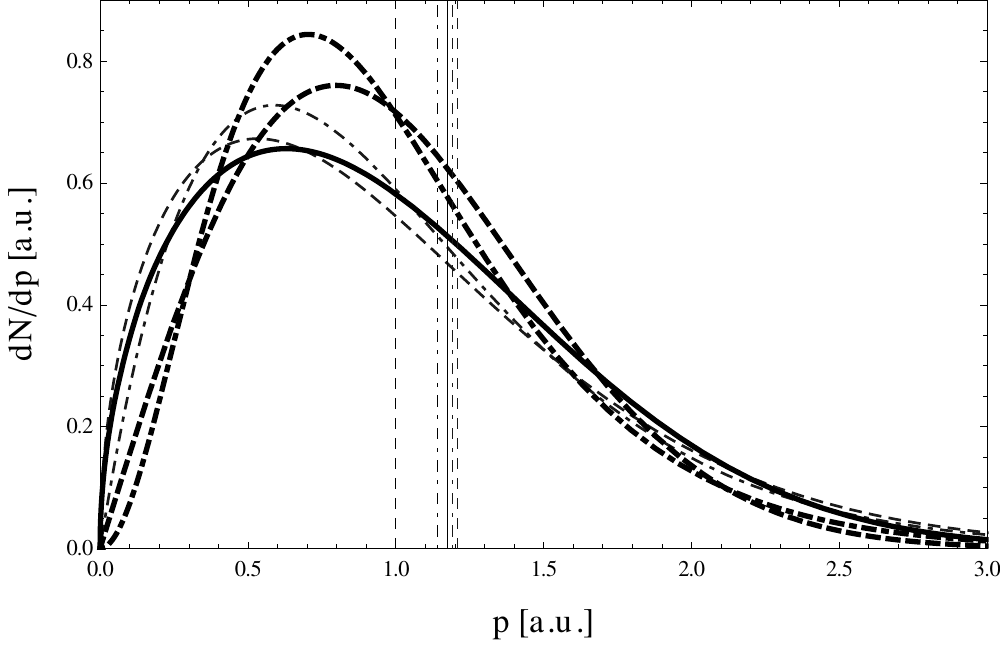} 
\caption{Momentum distribution functions with perfect mimicry in the first two cosmological observables corresponding to their first two moments and their next higher moment.
Depicted are distributions for the early decay (black, solid), the late decay (black, dashed), Pauli-Dirac (black, dash-dotted), Einstein-Bose (grey, dash-dotted) and decay in a matter-dominated universe (grey, dashed). Zeroth and first moment have been set to unity. Vertical lines mark the corresponding second moments (= skewness). Units are arbitrary.
}
\label{fig:fp_overplot}
\end{figure}
The predicted deviation between \nus HDM and dpHDM (with $g_\text{hdm}=2$) in the normalised root-mean-squares of absolute momenta
is thus found by combining~\eqref{grmsdec} and~\eqref{grmsnus} as
\begin{equation}
 \g_\text{ld}^\text{rms}/\g_{\nus}^\text{rms} \simeq 0.0105 \, \frac{\D\Neff^\frac{3}{4}}{\O_\text{dec}^0 h^2} \simeq 0.989 \, \frac{\D\Neff^\frac{3}{4}}{\mnuseff/\text{eV}} \, .
\end{equation}
Under the crude assumption of Gaussian error propagation this implies for~\eqref{hdmsignal} a prediction,
\begin{equation}
 \g_\text{ld}^\text{rms}/\g_{\nus}^\text{rms} = 1.66 \pm 0.81 \; (\text{1-}\s)\, .
\end{equation}
Since the momentum distribution functions differ between early and late decay as can be seen in Fig.~\ref{fig:fp_overplot}, these numbers differ between these two cases. We find  $ \langle p_\text{ed} \rangle_\text{rms} \simeq 1.03 \times \langle p_\text{ld} \rangle_\text{rms}$ and thus
\begin{equation}
  \g_\text{ed}^\text{rms}/\g_{\nus}^\text{rms} = 1.72 \pm 0.84 \; (\text{1-}\s)\, .
\end{equation}
So there is a 66\% (72\%) difference in the mean for the late (early) decay but the uncertainty of the prediction is decisively larger. 
Obviously, the difference to \nus HDM is too small to be detected in an extended analysis that would have to try to determine $\g^\text{rms}$ as additional free parameter. 
Not to mention that the observations need to be sensitive to this effect.
We do not expect that current observations or observations in the near future will be able to distinguish dpHDM from \nus HDM in all cases. In this sense, the mimicry can be ''perfect for our limited observational capabilities.``\footnote{
This usage of the notion of mimicry is actually closest to mimicry in biology.
}

It is tempting to look for a first hint if this negative conclusion could change in the next decade. With full Planck data available the galaxy survey EUCLID (to be launched 2020) will increase sensitivities dramatically by roughly an order of magnitude, cf.~\cite{Hamann:2012fe}.  Including galaxy surveys, eBOSS and DESI, and a new (Stage-IV) CMB polarimeter,
expected sensitivities go down to $\s(\sumnu) = 16 \meV$ and $\s(\Neff)= 0.020$~\cite{Abazajian:2013oma}.
These sensitivities would imply 1-$\s$ errors in the prediction of $\g_\text{dec}^\text{rms}/\g_{\nus}^\text{rms}$ that are nine times smaller than the predicted deviation for the current mean values~\eqref{hdmsignal}. However, these means can and will change, but for means half the size the 1-$\s$ errors were still $3.5$ times smaller than the predicted deviation.
If not from testing our standard interpretation alone, the opportunity to distinguish $\nus$HDM from dpHDM, motivates to study which future observations can how far probe the root-mean-square of absolute momenta of a free-streaming population. 
\section{Summary and Conclusions}
\label{sec:conclusions}
\begin{table}
 \begin{tabular}{l|c|c|c}
 \shortstack{daughter \\ masses} & \shortstack{mimicry in \\ 1st two}    & opportunity? & \nus\ equivalent  \\ \hline
$m_1 \ll m_2$ & \checkmark &  $2\times$larger $\Tnr$ & \shortstack{2\nus : $T_1=T_2$, \\ $m_2 >m_1=0$}   \\  \hline
$m_1=m_2$ & \checkmark & \shortstack{breaks down \\ in $\langle p \rangle_\text{rms}$} &  \shortstack{''perfect`` for \\ current obs.} \\ \hline  
$m_1\lesssim m_2$ & \checkmark & as \nus\ equivalent & \shortstack{2\nus : $T_1=T_2$, \\ $m_2 \gtrsim m_1>0$}   \\
\end{tabular}
\caption{Overview of different cases depending on the daughter masses. In every case there is mimicry in the first two cosmological observables. Given are opportunities to distinguish these cases from \nus HDM and how they are represented by thermal \nus\ populations.}
\label{tab:summary}
\end{table}
As summarised in Tab.~\ref{tab:summary}, it depends on the daughter masses where and how the mimicry breaks down. Every case can be implemented easily as \nus\ populations in existing numerical tools.
If the daughters do posses a large mass hierarchy, the temperature when dpHDM becomes non-relativistic is larger by a factor of two compared to \nus HDM, so that this case might be in stronger tension with CMB data. 
If the daughters are mass degenerate, they are indistinguishable from \nus HDM in analyses like~\cite{Hamann:2013iba}. 

Connecting cosmological ''observables`` with moments of the momentum distribution functions depicted in Fig.~\ref{fig:fp_overplot}, we find that for mass-degenerate daughters the mimicry breaks down only, if the next higher moment of the momentum distribution, the skewness, is considered. We define the \textit{normalised root-mean-square of absolute momenta} and find sizeable differences in its predicted value between dpHDM and \nus HDM. While these are certainly too small for current observations, this is a \textit{qualitative difference} compared to the attempt to distinguish different \nus HDM models, where the mimicry is perfect to arbitrary order of moments.

Other opportunities depend on the time of decay: For certain times during the BBN era, dpHDM becomes non-relativistic later than \nus HDM. A decay after BBN increases $\Neff|_\text{cmb} > \Neff|_\text{bbn}= \Neff^\text{sm}$, which motivates analyses that drop the BBN consistency relation.

To conclude, current cosmological observations are sensitive to sub-eV, not fully-thermalised HDM characterised by $\D\Neff < 1$ and $\mnuseff < \eV$. 
In that case, from a cosmological point of view \nus HDM is not preferred over dpHDM, neither from theoretical nor practical simplicity.
Our (current) blindness for the case of mass-degenerate daughters, should prevent us from premature conclusions when interpreting signals like~\eqref{hdmsignal}. Fortunately, there are various cases that can be considered easily and gainfully in likelihood analyses utilising available data already. 
After our proof of principle that, in contrast to different, thermal \nus HDM models, the mimicry of dpHDM breaks down at least in the root-mean-square of absolute momenta, it is an open question which observation due to which effect will be able to distinguish the two possibilities.

\subsection*{Acknowledgements}
\noindent
I would like to thank Jan Hamann for valuable discussions.
I acknowledge support from the German Academy of Science through the Leopoldina Fellowship Programme 
grant LPDS 2012-14.
 \begin{appendix}
\section{Free-streaming scale and transition time}
\label{appendix:A}
Starting from (93) in~\cite{Lesgourgues:2006nd},
\begin{equation}
 k^\text{fs}(T) =\left(\frac{3}{2} \frac{H^2(T) a^2(T)}{\langle v \rangle(T)}\right)^2 \text{, } \l^\text{fs}(T)=2\pi\frac{a(T)}{k^\text{fs}(T)},
\end{equation}
we approximate the average absolute velocity of the population $\langle v \rangle = \langle p\rangle/m$ at the transition, $\langle p \rangle =m$, as the speed of light $c$. We assume that it then decreases as $a^{-1} \propto T$, so that $\langle v \rangle(T) = c \, T/\Tnr $, if $\Tnr$ denotes the temperature of the Universe at the transition.\footnote{
As a side note,
we could also calculate $\langle v \rangle =\langle E_\text{kin} \rangle /m = \rho_\text{kin}/ (mn)$
and insert appropriate expressions for the kinetic energy density $\r_\text{kin}$.
}
The scale factor can be expressed in temperatures as $a(T)/a_0=T_0/T$ and $H^2 = H_0^2 \left(\O_\text{m} (T/T_0)^3 +\O_\L\right)$.
Inserting yields 
\begin{equation}
k^\text{fs}(T) \simeq 4.08 \times 10^{-4} \left(\O_\text{m} \fb{T}{T_0}{3} +\O_\L \right)^\frac{1}{2} \frac{T_0 \Tnr}{T^2} \frac{h}{\text{Mpc}}
\end{equation}
and
\begin{equation}
 \lfs(T) \simeq 1.54 \times 10^4 \left(\O_\text{m} \fb{T}{T_0}{3} +\O_\L \right)^{-\frac{1}{2}} \frac{T}{\Tnr} \frac{\text{Mpc}}{h} .
\end{equation}
For $k^\text{nr}=k^\text{fs}(T^\text{nr})$ we obtain~\eqref{knr}. 
For reference we provide the often used free-streaming scale of the population today,
\begin{equation}
 \lfs \simeq 1.54 \times 10^4 \frac{T_0}{\Tnr} h^{-1} \Mpc.
\end{equation}
Consistently we see from (93) of~\cite{Lesgourgues:2006nd}
\begin{equation}
\lfs(\tnr) =2 \pi (2/3)^{1/2} \langle v(\tnr) \rangle/H(\tnr)=\sqrt{6} \pi \tnr \, ,
\end{equation}
where we used $\langle v(\tnr) \rangle =1$ and $H=2/(3 t)$ in matter domination. We calculate the time when a population becomes non-relativistic in a Universe filled
with radiation and matter as
$
\tnr = \frac{\teq}{2-\sqrt{2}} \left( \left(\frac{\Teq}{\Tnr} -2 \right) \left( \frac{\Teq}{\Tnr} +1 \right)^{1/2}   +2 \right)
$,
where $\Teq$ $(\teq)$ denotes the temperature (time) at matter-radiation equality.

\section{Neutrino distribution specified by its moments}
\label{appendix:nudistr}
This appendix shall improve the accessibility of Sec.~\ref{sec:breakdown} for the reader.
We repeat findings of Cuoco, Lesgourgues, Mangano and Pastor in~\cite{Cuoco:2005qr} and refer the reader to
the original work.

The solution to the collisionless kinetic equations in a Lemaitre-Friedman-Robertson-Walker
universe is a Fermi-Dirac function $f(\vec{p}) = 1/(e^{(E-\m)/T}+1)$ with particle energy $E^2=|\vec{p}|^2 +m^2= p^2+m^2$.
We are interested in the standard situation with 
vanishing chemical potential $\m$ and early times, when neutrinos are ultra-relativistic.
Considering a radiation dominated universe, where $T\propto a^{-1}$, and defining the comoving momentum $y=pa$
one finds
\begin{equation}
\label{nudistr}
df(p,T_\nu)= \frac{1}{\pi^2} T_\nu^3 \frac{y^2}{e^y+1} dy \, ,
\end{equation}
where isotropy has been exploited to reduce the dimension of the differential and
which is normalised such that the integral yields the number density as required.
One can define a set of moments
\begin{equation}
\label{numoments}
Q^{(n)}_\nu= \frac{1}{\pi^2} \fb{4}{11}{\frac{3+n}{3}} T^{3+n} \int y^{2+n} f(y) dy
\end{equation}
in terms of the neutrino temperature $T_\nu= (4/11)^{1/3} T$. These moments can
specify the neutrino distribution $df(y)$ regardless of the specific case at hand.
In~\cite{Cuoco:2005qr} this is used to explore observation prospects for non-thermal contributions to the
standard neutrino spectrum. 
If the distribution decays at large comoving
momentum as $e^{-y}$, it admits moments of all orders.
In~\cite{Cuoco:2005qr} the neutrino distribution is given in terms of its moments as
\begin{equation}
df(y)= \frac{y^2}{e^y+1} \sum_{m=0}^\infty \sum_{k=0}^m
c_k^{(m)} Q^{(k)}_\nu T_\nu^{-k} P_m(y) dy
\end{equation}
with $P_m(y)= \sum_{k=0}^m c_k^{(m)} y^k$, $m$ being the degree of $P_m(y)$ and $c_k^{(m)}$ being a coefficient,
denoting the set of polynomials orthonormal with respect to the measure $y^2/(e^y+1)$, i.e.,
$\int_0^\infty dy \frac{y^2}{e^y+1} P_n(y)P_m(y) =\d_{nm}$.
For a Fermi-Dirac distribution all moments can be expressed in terms of the lowest moment $Q^{(0)}= n_\nu$ or,
equivalently, as functions of the temperature $T_\nu$ since it is the only independent parameter.
For neutrinos the (first two) cosmological observables can be written as
\begin{align}
 \Neff = \frac{120}{7 \pi^2} T_\nu^{-4} \sum_\a Q_{\nu \a}^{(1)}
\end{align}
and
\begin{align}
\O_\nu^0 &= \frac{\sumnu}{\rhoc} Q_\nu^{(0)} \fb{T_0}{T}{3}  \nonumber \\ 
&\Leftrightarrow \O_\nu^0 h^2 \simeq 0.162 \, \frac{\sumnu}{\text{eV}} T^{-3} Q_\nu^{(0)} ,
\end{align} 
where it is assumed that the three neutrinos share the same distribution today and
the small correction from $e^+e^-$ annihilation,  $\Neff^\text{sm}=3.046 \neq 3$, is incorporated in the numerical prefactor.
In~\cite{Cuoco:2005qr} $\Neff$ and $f_\nu$ are used to probe deviations from the standard neutrino spectrum
 in the first two moments.

Also the following facts are used in our discussion:
The average absolute neutrino momentum $\langle p_\nu \rangle$ expressed by moments of their distribution~\eqref{numoments} reads
\begin{equation}
 \label{averagenumomentum} 
\langle p_\nu \rangle = \frac{Q_\nu^{(1)}}{Q_\nu^{(0)}} 
= \frac{7 \pi^4}{180 \zeta(3)} \fb{4}{11}{\frac{1}{3}} T \simeq 3.15 \, T_\nu
\end{equation}
 and the root-mean-square of absolute momenta $\langle p_\nu \rangle_\text{rms} =(Q_\nu^{(2)}/Q_\nu^{(0)})^{1/2}$ can be calculated analogously as
\begin{equation}
 \label{rmsnumomentum}
\langle p_\nu \rangle_\text{rms}
= \fb{15 \zeta(5)}{\zeta(3)}{\frac{1}{2}} \fb{4}{11}{\frac{1}{3}} T \simeq 3.6 \, T_\nu \, .
\end{equation}

 \end{appendix}

\phantomsection 
\addcontentsline{toc}{chapter}{References}
\bibliography{mimicry}

\end{document}

%% file: commands.tex
\def\a{\alpha}

\def\d{\delta}

\def\g{\gamma}

\def\l{\lambda}
\def\m{\mu}
\def\n{\nu}
\def\o{\omega}

\def\r{\rho}
\def\s{\sigma}
\def\t{\tau}

\def\z{\zeta}
\def\D{\Delta}

\def\G{\Gamma}

\def\L{\Lambda}
\def\O{\Omega}

\newcommand{\meV}{\text{ meV}}
\newcommand{\eV}{\text{ eV}}

\newcommand{\MeV}{\text{ MeV}}

\newcommand{\Mpc}{\text{ Mpc}}

\newcommand{\seconds}{\text{ s}}

\newcommand{\nus}{\ensuremath{\nu_s}}

\newcommand{\mnus}{m_{\nu_s}}
\newcommand{\mnuseff}{m_\text{hdm}^\text{eff}}

\newcommand{\Trh}{T_\text{rh}}

\newcommand{\rhoc}{\r_{\text{c}}}
\newcommand{\teq}{{t_\text{eq}}}
\newcommand{\Teq}{T_\text{eq}}

\newcommand{\Neff}{N_\text{eff}}
\newcommand{\sumnu}{\sum m_\nu}

\newcommand{\lfs}{\l^\text{fs}}

\newcommand{\gast}{g_\ast}

\newcommand{\gastt}{g_\ast^0}

\newcommand{\gasts}{g_{\ast s}}
\newcommand{\gastsd}{g_{\ast s}^\text{d}}

\newcommand{\gastst}{g_{\ast s}^0}

\newcommand{\rhorad}{\rho_\text{rad}}

\newcommand{\Tdec}{T_\text{d}}

\newcommand{\Tnr}{T^\text{nr}}
\newcommand{\tnr}{t^\text{nr}}

\newcommand{\Tnus}{T_{\nu_s}}

\newcommand{\pini}{p_\text{ini}}
\newcommand{\nini}{n_\text{ini}}

\newcommand{\fb}[3]{\left(\frac{#1}{#2}\right)^{#3}}
\newcommand{\order}[1]{\ensuremath{\mathcal{O}\left(#1\right)}}